\documentclass[10pt,conference]{IEEEtran}
\usepackage{graphicx,cite,amssymb,amsmath,psfrag,subfigure}
\usepackage{graphicx}
\usepackage{amssymb}
\usepackage{amsmath}
\usepackage{cite}
\usepackage{subfigure}
\usepackage{mathrsfs}
\usepackage[displaymath,mathlines]{lineno}
\usepackage{color}
\usepackage{tabulary}
\usepackage{multirow}

\newtheorem{theorem}{Theorem}

\newtheorem{proposition}{Proposition}

\newtheorem{remark}{Remark}

\DeclareMathOperator{\E}{\mathbb{E}}

\DeclareMathOperator{\var}{\mathbb{V}\mathrm{ar}}

\newcommand{\EX}[1]{\E\left\{{#1}\right\}}

\newcommand{\varx}[1]{\var\left\{{#1}\right\}}

\newcommand{\CG}[2]{\mathcal{CN}\left({#1},{#2}\right)}

\newcommand{\B}[1]{{\mathbf{#1}}}

\newcommand{\Pp}{\rho_{\mathrm{p}}}

\newcommand{\Pd}{\rho_{\mathrm{d}}}

\makeatletter
\def\@setsize#1#2#3#4{
    \@nomath#1
    \let\@currsize#1
    \baselineskip #2
    \baselineskip \baselinestretch\baselineskip
    \parskip \baselinestretch\parskip
    \setbox\strutbox \hbox{
        \vrule height.7\baselineskip
            depth.3\baselineskip
            width\z@}
    \skip\footins \baselinestretch\skip\footins
    \normalbaselineskip\baselineskip#3#4}
\makeatother

\makeatletter
\newcommand{\setstretch}[1]{
    \def\baselinestretch{#1}%
    \@currsize
    }
\makeatother



\def\BibTeX{{\rm B\kern-.05em{\sc i\kern-.025em b}\kern-.08em
    T\kern-.1667em\lower.7ex\hbox{E}\kern-.125emX}}

\newcounter{eqncnt}

\newcounter{eqnback}
\begin{document}
\title{
    Cell-Free Massive MIMO: \\ Uniformly Great Service For Everyone}
\author{\IEEEauthorblockN{Hien Quoc Ngo\IEEEauthorrefmark{1}, Alexei Ashikhmin\IEEEauthorrefmark{2}, Hong Yang\IEEEauthorrefmark{2}, Erik
G. Larsson\IEEEauthorrefmark{1}, and Thomas L.
Marzetta\IEEEauthorrefmark{2}}
\IEEEauthorblockA{\IEEEauthorrefmark{1}Department of Electrical
Engineering (ISY), Link\"{o}ping University, 581 83 Link\"{o}ping,
Sweden} \IEEEauthorblockA{\IEEEauthorrefmark{2}Bell Laboratories,
Alcatel-Lucent, Murray Hill, NJ 07974, USA\vspace{-1cm}}
\thanks{
The work of H.~Q.\ Ngo and E.~G.\ Larsson was supported  in part
by the Swedish Research Council (VR), and ELLIIT. } }

%
%

\maketitle

\begin{abstract}
We consider the downlink of Cell-Free Massive MIMO systems, where
a very large number of distributed access points (APs)
simultaneously serve a much smaller number of users. Each AP uses
local channel estimates obtained from received uplink pilots and applies
conjugate beamforming to transmit data to the users. We derive a
closed-form expression for the achievable rate. This
expression enables us to design an optimal max-min power control
scheme that gives  equal quality of service to all users.

 We further compare the performance of the Cell-Free Massive MIMO system to that
of a conventional small-cell network and show that the throughput
of the Cell-Free system is much more concentrated around its
median compared to that of the small-cell system. The Cell-Free
Massive MIMO system can provide an almost $20-$fold increase in  95\%-likely per-user throughput, compared with the small-cell
system. Furthermore, Cell-Free systems are  more
robust to   shadow fading correlation than  small-cell systems.
\end{abstract}


\section{Introduction} \label{Sec:Introduction}

In Massive MIMO, large collocated or distributed antenna arrays are
deployed at wireless base stations
\cite{Eri:13:MCOM}. Collocated Massive MIMO architectures, where
all service antennas are located in a compact area, have the
advantage that the backhaul requirements are low. By contrast, in
distributed Massive MIMO, the service antennas are spread out over
a wide area. Owing to its ability to fully exploit  macro-diversity and
differences in path loss, distributed Massive MIMO
 can potentially offer a very high probability of coverage.
 At the same time, it can offer all the advantages of collocated Massive MIMO.
 Despite its potential, however, fairly little work on distributed Massive MIMO
is available  in  the  literature.
Some comparisons between distributed and collocated systems were
performed in \cite{TH:13:ACSSC} for the uplink and in
\cite{LD:14:COM,HWA:14:JSTSP} for the downlink.  In
\cite{LD:14:COM,HWA:14:JSTSP}, perfect channel state information
was assumed to be available  both at the base station and at the users.

On a parallel avenue, small-cell systems, where the base stations  do not
cooperate coherently, are often viewed as enablers for high  spectral efficiency and  coverage
\cite{LHYS:13:WCNC}. However, a comparison between small-cell and
distributed Massive MIMO is not yet available.

\textbf{Contribution of the paper:}
We consider a distributed Massive MIMO system where
a large number of service antennas, called access points
(APs), and a much smaller number of autonomous users are
distributed at random over a wide area.  All APs   cooperate via
a backhaul network, and serve all users in the same time-frequency
resource  via time-division duplex (TDD) operation. There are   no
cells or cell boundaries. Hence, we call this system
``Cell-Free Massive MIMO''.
Cell-Free Massive MIMO   is expected to offer many advantages: 1) a huge throughput, coverage probability and
energy efficiency; and 2) averaging out of small-scale fading and
uncorrelated noise so that the performance  depends only on the
large-scale fading. In this paper, we restrict the discussion to the
downlink of Cell-Free systems with conjugate beamforming. We
assume that the channels are estimated through uplink pilots. The
paper makes the following specific contributions:
\begin{itemize}
\item We derive a closed-form expression for the achievable rate
with a finite number of APs, taking into account channel
estimation errors, power control, and the non-orthogonality of
pilot sequences. In \cite{YM:13:ACCCC},  corresponding capacity
  bounds  were derived assuming  pilot
sequences are  orthogonal.

\item We design two pilot assignment schemes: \emph{random pilot
assignment} and \emph{greedy pilot assignment}. We show that the
greedy pilot assignment is   better.

\item We propose a max-min power   control algorithm that   maximizes the
smallest of all user rates, under a
per-AP power constraint. This   power control problem is formulated as
 a quasi-convex optimization problem.

\item We compare the performance of Cell-Free Massive MIMO and
conventional small-cell systems under uncorrelated and correlated
shadowing models. Conclusions of this comparison are given in Section~\ref{Sec:Conclusion}.
\end{itemize}


\section{System Model} \label{Sec:SysModel}

We consider a network of $M$ APs that serve $K$ users in the same
time-frequency resource. All APs and users are equipped with a
single antenna, and they are randomly located in a large area. In
this work, we consider the downlink with conjugate beamforming. We
consider only the conjugate beamforming technique since it is
simple and can be implemented in a distributed manner. The
transmission from the APs to the users proceed by TDD operation,
including two phases for each coherence interval: uplink training
and downlink payload data transmission. In the uplink training
phase, the users send pilot sequences to the APs and each AP
estimates its own CSI. This is performed in a decentralized
fashion. The so-obtained channel estimates are used to precode the
transmit signals.

\subsection{Uplink Training}

Let $T$ be the length of the coherence interval (in samples), and
let $\tau$ be the length of uplink training duration (in samples) per
coherence interval. It is required that $\tau<T$. During the
training phase, all $K$ users simultaneously send pilot sequences
of length $\tau$ samples to the APs. Let $\pmb{\varphi}_k \in
\mathbb{C}^{\tau\times 1}$, where $\|\pmb{\varphi}_k\|^2=1$, be
the pilot sequence used by the $k$th user. Then, the $\tau\times
1$  pilot vector received at the $m$th AP is
\begin{align}\label{eq:pilot1}
\B{y}_m = \sqrt{\tau \Pp}\sum_{k=1}^K g_{mk} \pmb{\varphi}_k +
\B{w}_m,
\end{align}
where $\Pp$ is the normalized transmit signal-to-noise ratio (SNR)
of each pilot symbol, $\B{w}_m$ is a vector of  additive noise  at
the $m$th AP---whose elements are i.i.d.\ $\CG{0}{1}$ RVs, and
$g_{mk}$ denotes the channel coefficient between the $k$th user
and the $m$th AP. The channel $g_{mk}$ is modeled as follows:
\begin{align}\label{eq:gmk}
g_{mk} =  \beta_{mk}^{1/2}h_{mk},
\end{align}
where $h_{mk}$ represents the small-scale fading, and $\beta_{mk}$
represents the path loss and large-scale fading. We assume that
$h_{mk}$, $m=1, \ldots, M$, $k=1, \ldots K$, are i.i.d.\
$\CG{0}{1}$ RVs. The assumption of independent small-scale fading
is reasonable since the APs and the users are distributed over a
wide area, and hence, the set of scatterers is likely to be
different for each AP and each user.

Based on the received pilot signal $\B{y}_m$, the $m$th AP will
estimate the channel $g_{mk}, \forall k=1, ..., K$. Let
$\tilde{y}_{mk} \triangleq \pmb{\varphi}_k^H\B{y}_m$ be the
projection of $\B{y}_m$ on $\pmb{\varphi}_k^H$:
\begin{align}\label{eq:tidley1}
\tilde{y}_{mk}  = \sqrt{\tau \Pp} g_{mk}  + \sqrt{\tau
\Pp}\sum_{k'\neq k}^K g_{mk'} \pmb{\varphi}_k^H \pmb{\varphi}_{k'}
+ \pmb{\varphi}_k^H\B{w}_m.
\end{align}
Although, for arbitrary pilot sequences, $\tilde{y}_{mk}$ is not a
sufficient statistic for the estimation of $g_{mk}$, one can still
use  $\tilde{y}_{mk}$ to obtain suboptimal estimates. In the case
where any pair of pilot sequences is either fully correlated or
exactly orthogonal, then $\tilde{y}_{mk}$ is a sufficient
statistic, and  estimates based on $\tilde{y}_{mk}$ are optimal.
The MMSE estimate of $g_{mk}$ given $\tilde{y}_{mk}$ is
\begin{align}\label{eq:MMSE est1}
\hat{g}_{mk}
=\frac{\sqrt{\tau\Pp}\beta_{mk}\tilde{y}_{mk}}{\tau\Pp\sum_{k'=1}^K\beta_{mk'}\left|\pmb{\varphi}_k^H
\pmb{\varphi}_{k'}\right|^2+1}.
\end{align}

\begin{remark} \label{remark pilot}
If $\tau \geq K$, then we can choose $\pmb{\varphi}_1, \cdots,
\pmb{\varphi}_K$ so that they are pairwisely orthogonal, and
hence, the second term in \eqref{eq:tidley1} disappears.  However,
owing to the limited length of the coherence interval, in general,
$\tau < K$, and non-orthogonal pilot sequences must be used by
different users.  The channel estimate $\hat{g}_{mk}$ is degraded
by pilot signals transmitted from other users (the second term in
\eqref{eq:tidley1}). This causes the so-called pilot contamination
effect. There is still room for a considerable number of
orthogonal pilots. For the case of pedestrians with mobility less
than 3 km/h, at a 2 GHz carrier frequency the product of the
coherence bandwidth and the coherence time is equal to $17,640$
(assuming a delay spread of $4.76$ $\mu$s) \cite{TV:05:Book}, so
if half of the coherence time is utilized for pilots then there is
an available pool of $8820$ mutually orthogonal pilots.
\end{remark}

\subsection{Downlink Payload Data Transmission}

The APs treat the channel estimates as the true channels, and use
  conjugate beamforming  to transmit signals to the $K$
users. Let $s_k$, $k=1, \ldots, K$, where $\EX{|s_k|^2}=1$, be the
symbol intended for the $k$th user. With conjugate beamforming,
the transmitted signal from the $m$th AP is
\begin{align}\label{eq:xm}
x_m = \sqrt{\Pd}\sum_{k=1}^K \eta_{mk}^{1/2}\hat{g}_{mk}^\ast s_k,
\end{align}
where  $\eta_{mk}$, $m=1, \ldots, M$, $k=1, \ldots K$, are
power control coefficients chosen to satisfy the following average power
constraint at each AP:
\begin{align}
\EX{|x_m|^2}\leq \Pd.
\end{align}
  With the channel
model in \eqref{eq:gmk}, the power constraint $\EX{|x_m|^2}\leq
\Pd$ can be rewritten as:
\begin{align}\label{eq:pct}
\sum_{k=1}^K \eta_{mk}\gamma_{mk} \leq 1,
\end{align}
where
\begin{align}\label{eq:gamma1}
\gamma_{mk}\triangleq \EX{\left|\hat{g}_{mk} \right|^2} =
\frac{{\tau\Pp}\beta_{mk}^2}{\tau\Pp\sum_{k'=1}^K\beta_{mk'}\left|\pmb{\varphi}_k^H
\pmb{\varphi}_{k'}\right|^2+1}.
\end{align}

The received signal at the $k$th user is given by
\begin{align}\label{eq:rk1}
r_k \!\!=\!\!\! \sum_{m=1}^M\!\! g_{mk}x_m \!+\! n_k \!=\!
\sqrt{\Pd}\!\!\sum_{m=1}^M\!\sum_{k'=1}^K
\!\!\eta_{mk'}^{1/2}g_{mk}\hat{g}_{mk'}^\ast s_{k'} \!+\! n_k,
\end{align}
where $n_k$ represents additive Gaussian noise at the $k$th user. We
assume that $n_k \sim \CG{0}{1}$. Then $s_k$ will be detected from
$r_k$.

\setcounter{eqnback}{\value{equation}} \setcounter{equation}{11}
\begin{figure*}[!t]
\begin{align}\label{eq:Theo_rateexpr1}
R_k^{\text{cf}}
 =
 \log_2
    \left(
    1 + \frac{\Pd \left(\sum_{m=1}^M \eta_{mk}^{1/2}\gamma_{mk} \right)^2 }{ \Pd\sum_{k'\neq k}^K\left(\sum_{m=1}^M \eta_{mk'}^{1/2}\gamma_{mk'}\frac{\beta_{mk}}{\beta_{mk'}} \right)^2| \pmb{\varphi}_{k'}^H\pmb{\varphi}_{k}|^2 + \Pd\sum_{k'=1}^K\sum_{m=1}^M \eta_{mk'}\gamma_{mk'}\beta_{mk} +1 }
    \right).
\end{align}
\hrulefill
\end{figure*}
\setcounter{eqncnt}{\value{equation}}
\setcounter{equation}{\value{eqnback}}

\section{Achievable Rate}

In this section, we derive a closed-form expression for the
achievable rate, using similar analysis techniques as in
\cite{HH:03:IT}. We assume that each user has knowledge of the channel
statistics but not of the channel realizations. The received signal $r_k$ in \eqref{eq:rk1} can be
rewritten as
\begin{align}\label{eq:rate1} &r_k \!=\!
  {\tt DS}_k\cdot
  s_{k} + {\tt EN}_k,
\end{align}
where ${\tt DS}_k \triangleq  \sqrt{\Pd}\EX{\sum_{m=1}^M
\eta_{mk}^{1/2}g_{mk}\hat{g}_{mk}^\ast}$ is a deterministic factor that scales the
desired
signal, and ${\tt EN}_k$ is an ``effective noise'' term which equals
the RHS of \eqref{eq:rk1} minus ${\tt DS}_k\cdot
  s_{k}$. A straightforward
calculation shows that the effective noise and the desired signal
are uncorrelated. Therefore, by using the fact that uncorrelated
Gaussian noise represents the worst case, we obtain the following
achievable rate of the $k$th user for Cell-Free (cf) operation:
\begin{align}\label{eq:rateexpr1}
R_k^{\text{cf}}
 =
 \log_2
    \left(
    1 + \frac{\left|{\tt DS}_k \right|^2}{ \varx{{\tt EN}_k} }
    \right).
\end{align}

We next provide a new exact closed-form expression for the
achievable rate given by \eqref{eq:rateexpr1}, for a finite number
of APs. (The proof uses a standard Gaussian-is-worst-case-noise
argument and is omitted here due to space limitations.)

\begin{theorem}\label{theorem rate}
An achievable rate of the transmission from the APs to the $k$th
user    is given by
\eqref{eq:Theo_rateexpr1}, shown at the top of the next page.
\end{theorem}
\setcounter{equation}{12}

\section{Pilot Assignment and Power Control}

In this section, we will present  methods to  assign the pilot
sequences to the users, and to control the power. Note
that pilot assignment and power control are decoupled problems
because the pilots are not power controlled.

\subsection{Pilot Assignment}

As discussed in Remark~\ref{remark pilot}, non-orthogonal pilot
sequences must be used by different users due to the limited
length of the coherence interval. The main question is how
the pilot sequences should be assigned to the $K$ users. In the
following, we present two simple pilot assignment schemes.

\subsubsection{Random Pilot Assignment}

Since the length of the pilot sequences is $\tau$, we can have
$\tau$ orthogonal pilot sequences. With random pilot assignment,
 each user will be randomly assigned one pilot sequence from
the set of $\tau$ orthogonal pilot sequences.

This random pilot assignment method is simple. However, with high
probability,  two users that are in close vicinity of each other
will use the same pilot sequence. As a result, the performance of
these users is not good, due to the high pilot contamination
effect.

\subsubsection{Greedy Pilot Assignment}

Next we propose a simple greedy pilot assignment method, that
addresses the shortcomings of random pilot assignment. With greedy
pilot assignment, the $K$ users are first   randomly assigned $K$
pilot sequences. Then the worst-user (with lowest rate), say  user
$k$,   updates its pilot sequence so that its pilot contamination
effect is minimized. The algorithms then proceeds iteratively.
Note that the greedy pilot assignment can be performed at a
central processing unit.

The pilot contamination effect at the $k$th user is
reflected by the second term in \eqref{eq:tidley1} which has
variance
$\sum_{k'\neq k}^K \beta_{mk'} \left|\pmb{\varphi}_k^H
\pmb{\varphi}_{k'} \right|^2.$
So, the worst-user (the $k$th user) will find a new pilot sequence
which minimizes the pilot contamination, summed over all APs:
 \begin{align}\label{eq GPA 2}
\arg\min_{\pmb{\varphi}_k} \sum_{m=1}^M\sum_{k'\neq k}^K
\beta_{mk'} \left|\pmb{\varphi}_k^H \pmb{\varphi}_{k'} \right|^2.
\end{align}
Since $\|\pmb{\varphi}_k\|^2=1$, the expression in \eqref{eq GPA 2} is
a Rayleigh quotient, and hence, the updated pilot sequence of
the $k$th user is the eigenvector which corresponds to the
smallest eigenvalue of the matrix $\sum_{m=1}^M\sum_{k'\neq k}^K
\beta_{mk'} \pmb{\varphi}_{k'}\pmb{\varphi}_{k'}^H$.

\subsection{Power Control}

 In our work, we focus on   max-min power control. Specifically,
given realizations of the large-scale fading, we find the  power control
coefficients $\eta_{mk}$, $m=1, \cdots, M, k=1, \cdots, K$, which
  maximize the minimum of the rates of   all users, under the power
constraint \eqref{eq:pct}:
\begin{align}\label{eq opt 1}
    \left.%
\begin{array}{l}
  \displaystyle\max_{\{\eta_{mk}\}} \hspace{0.8 cm} \min_{k=1, \cdots, K} R_k^{\text{cf}}  \\
  \text{subject to}
     \hspace{0.55 cm} \sum_{k=1}^K \eta_{mk}\gamma_{mk} \leq 1, ~  m=1, ..., M\\
     \hspace{2.0 cm}  \eta_{mk} \geq 0, ~ k=1, ..., K, ~  m=1, ..., M.\\
\end{array}%
\right.
\end{align}
Define   $\varsigma_{mk}\triangleq \eta_{mk}^{1/2}$. Then, by
introducing the slack variables $\varrho_{k'k}$ and
$\vartheta_{m}$, and using \eqref{eq:Theo_rateexpr1}, we can
reformulate \eqref{eq opt 1} as
\begin{align}\label{eq opt 2}
    \left.%
\begin{array}{l}
  \displaystyle\max_{\{\varsigma_{mk}, \varrho_{k'k},  \vartheta_{m}\}} \hspace{0.1 cm} \min_{k=1, \cdots, K}
    \frac{ \left(\sum_{m=1}^M \gamma_{mk}\varsigma_{mk} \right)^2 }{ \sum\limits_{k'\neq k}^K\!\!|\pmb{\varphi}_{k'}^H\pmb{\varphi}_{k}|^2 \varrho_{k'k}^2+\!\! \sum\limits_{m=1}^M\!\!\beta_{mk} \vartheta_{m}^2 +\frac{1}{\Pd} }  \\
  \text{subject to}
     \hspace{0.55 cm} \sum_{k'=1}^K \gamma_{mk'}\varsigma_{mk'}^2 \leq \vartheta_{m}^2 , ~ m=1, ..., M\\
     \hspace{2.0 cm}  \sum_{m=1}^M \gamma_{mk'}\frac{\beta_{mk}}{\beta_{mk'}}\varsigma_{mk'} \leq \varrho_{k'k}, ~ \forall k'\neq k\\
     \hspace{2.0 cm}  0 \leq \vartheta_{m} \leq 1, ~ m=1, ..., M\\
     \hspace{2.0 cm}  \varsigma_{mk} \geq 0, ~ k=1, ..., K, ~ m=1, ..., M.\\
\end{array}%
\right.
\end{align}

We then have the following result.  (The proof is
omitted due to space constraints.)
\begin{proposition}\label{prop1}
The optimization problem   \eqref{eq opt 2} is quasi-concave.
\end{proposition}

Since the optimization problem \eqref{eq opt 2} is quasi-concave,
it can be solved efficiently by using bisection and solving   a sequence of convex
feasibility problems \cite{BV:04:Book}.

\section{Numerical Results and Discussions}

We provide some numerical   results on Cell-Free Massive MIMO performance.
The $M$ APs and $K$ users
are uniformly distributed at random within a square of size
$1000\times 1000$ $\text{m}^2$. The coefficient $\beta_{mk}$
models the path loss and shadow fading, according to
\begin{align}\label{eq:betmk}
\beta_{mk} = \text{PL}_{mk}\cdot
10^{\frac{\sigma_{\text{sh}}z_{mk}}{10}},
\end{align}
where $\text{PL}_{mk}$ represents  path loss, and
$10^{\frac{\sigma_{\text{sh}}z_{mk}}{10}}$ represents  shadow
fading with  standard deviation $\sigma_{\text{sh}}$, and
$z_{mk}\sim\mathcal{N}(0,1)$. We use a three-slope model for the
path loss \cite{TSG:01:VTC}: the path loss exponent equals $3.5$
if the distance between the $m$th AP and the $k$th user (denoted by
$d_{mk}$) is greater than $d_1$, equals $2$ if $d_1\geq d_{mk}>
d_0$, and equals $0$ if $d_{mk} \leq d_0$. To determine the path loss in
absolute numbers, we  employ the Hata-COST231 propagation model
for  $d_{mk}>d_1$.


In all examples, we choose the following parameters:  the carrier
frequency is $1.9$ GHz, the AP radiated power is $200$ mW, the noise
figure (uplink and downlink) is $9$ dB, the AP antenna height is $15$ m,
the user antenna height is $1.65$ m, $\sigma_{\text{sh}}=8$ dB,
$d_1=50$, and $d_0=10$ m. To avoid   boundary effects, and to
imitate a network with infinite area, the square area is wrapped
around.


\begin{figure}[t!]
\centerline{\includegraphics[width=0.4\textwidth]{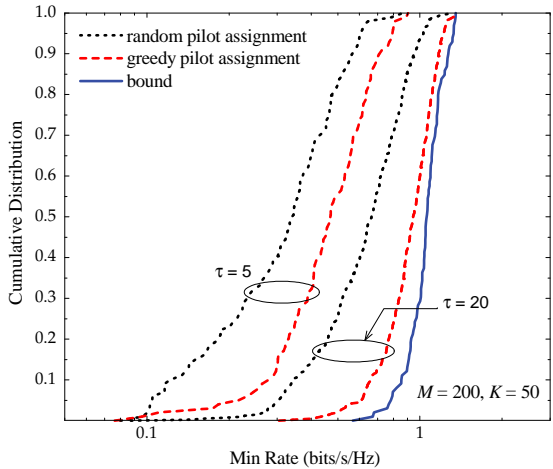}}
\caption{The minimum per-user rate for different $\tau$, without
power control. Here, $M=200$, $K=50$, and $D=1$ km.\label{fig:2}}
\end{figure}

\subsection{Pilot Assignment}

We first examine the performance of Cell-Free Massive MIMO with
different pilot assignment schemes, assuming that no power control
is performed. Without power control, all APs transmit with full
power, $\eta_{mk}=\eta_{mk'}$, $\forall k=1, \ldots, K$.
Figure~\ref{fig:2} shows the cumulative distribution of the
minimum per-user rate which is given in \eqref{eq:Theo_rateexpr1}
(the pre-log penalty  due to the training phase is not accounted
here), for $M=200$, and $K=50$, and for different $\tau$. The
``bound'' curve shows the case where all $K$ users are assigned
orthogonal pilot sequences, so there is no pilot contamination. As
expected, when $\tau$ increases, the pilot contamination effect
reduces, and hence, the rate increases. We can see that  when
$\tau=20$, then by using greedy pilot assignment
 the 95\%-likely minimum rate can be doubled as compared to when random pilot
assignment is used. In addition, even when $\tau$ is very small
($\tau=5$),
 Cell-Free Massive MIMO still provides good service for all
users. More importantly, the   gap between  the performance of
greedy pilot assignment and the  bound  is not significant. Hence,
the greedy pilot assignment scheme is  fairly good and henceforth
this is the method we will use. 

\begin{figure}[t!]
\centerline{\includegraphics[width=0.4\textwidth]{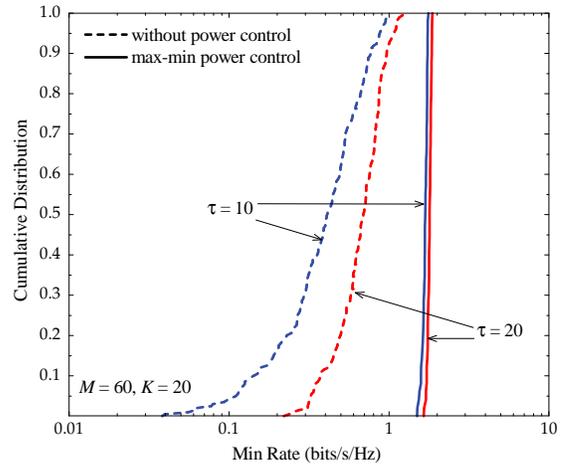}}
\caption{The minimum rates without power control and with max-min
power control. Here, $M=60$, $K=20$, and $\tau=10$ and
20.\label{fig:3}}
\end{figure}

\subsection{Max-Min Power Control}

In the following, we will examine effectiveness of the
max-min power control. Fig.~\ref{fig:3} shows the the cumulative
distribution of the achievable rates with max-min and no power
control respectively, for $M=60$, $K=20$, and $\tau=10, 20$.  We can
see that with max-min power control, the system performance
improves significantly. When $\tau=10$, the max-min power allocation
can improve the 95\%-likely rate by a factor of $15$ compared to
the case of without power control. In addition, by using power
control, the effect of pilot contamination can be  notably reduced.

\subsection{Cell-Free Massive MIMO versus Small-Cell Systems}

Next we compare the performance of Cell-Free Massive MIMO to that of
small-cell systems.

\subsubsection{Small-Cell Systems}

To model a  small-cell system, we assume that each user is served
by exactly one AP. For a given user, the available AP with the
largest average received power is selected. Once   an AP is chosen
by some user, this AP then becomes unavailable. The AP selection
is done user by user in a random order. Mathematically, we let
$m_k$ be the AP chosen by the $k$th user:
$$m_k\triangleq \arg\!\!\!\!\!\!\!\!\!\!\!\!\max_{m\in\{\text{available APs}\}}\beta_{mk}.$$

In small-cell operation, the users estimate the channels. Let
$\pmb{\varphi}_{k} \in \mathbb{C}^{\tau\times 1}$ be the pilot
sequence transmitted from the $m_k$th AP, and let $\Pp$ be the
average transmit power per pilot symbol. We assume that the APs
transmit with full power. Similarly to the derivation of Cell-Free
Massive MIMO (details omitted due to space constraints), we can
derive an achievable rate for  the $k$th user in a small-cell
system as:
\begin{align*}
    R_k^{\mathrm{sc}}
    &=
    \EX{\log_2\!\!\!
        \left(\!\!
        1\!+\! \frac{\Pd\left|\hat{g}_{{m_k}k}\right|^2}{\Pd(\beta_{m_kk} \!-\! \mu_{{m_k}k})+\Pd\!\!\sum\limits_{k'\neq k}^K \!\beta_{{m_{k'}}k} + 1}
        \!\!\right)},
\end{align*}
where $\hat{g}_{{m_k}k}\sim \CG{0}{\mu_{{m_k}k}}$ is the channel
estimate of ${g}_{{m_k}k}$, and $\mu_{{m_k}k} \triangleq
\sqrt{\tau\Pp}\beta_{{m_k}k}^2/(\tau\Pp\sum_{k'=1}^K\beta_{{m_{k'}}k}\left|\pmb{\varphi}_k^H
\pmb{\varphi}_{k'}\right|^2+1)$.

\begin{figure}[t!]
\centerline{\includegraphics[width=0.4\textwidth]{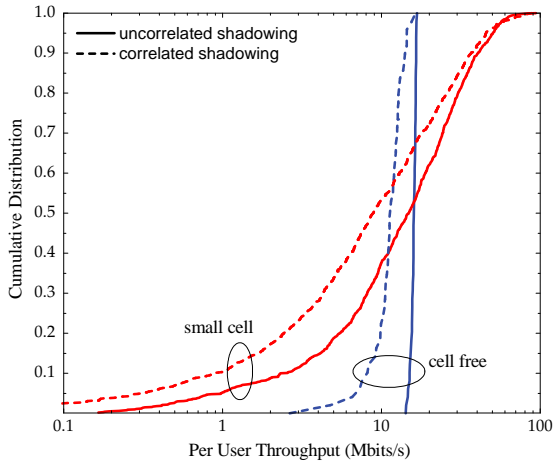}}
\caption{The cumulative distribution of the throughput per user
for correlated and uncorrelated shadow fading. Here, $M=60$,
$K=20$, and $\tau=10$. \label{fig:5}}
\end{figure}

\subsubsection{Spatial Shadowing Correlation Models}

Transmitters/receivers that are in close vicinity
of each other will experience correlated
shadow fading. Next, we  investigate the
effect of shadowing correlation on both small-cell and Cell-Free Massive MIMO
systems.

Most  existing correlation models for the shadow fading
model two correlation effects from the user perspective:
cross-correlation and spatial correlation \cite{WTN:08:VT}. The
cross-correlation effect represents the correlation between the
shadowing coefficients from different base stations to a given
user. The spatial correlation effect represents the correlation
due to the relative positions between the users. This model
neglects the effects of cross-correlation and spatial correlation
from the base station sides. Doing so is reasonable when the base
station antennas are located high above the ground, and all base
stations are well separated. However, in our system models, both
the APs and the users are located randomly in the network, and
they may be close to each other. In addition, the AP antenna
elevation is not very high. Hence, the correlation at the AP
side should be taken into account as well. We will use the following
modified correlation model for the shadow fading:
\begin{align*}
z_{mk}\!=\!\sqrt{\rho_1}a_m \!+\! \sqrt{1\!-\!\rho_1}b_k, ~
m\!=\!1,\ldots, M, ~ k\!=\!1,\ldots, K,
\end{align*}
where $a_m\sim\mathcal{N}(0,1)$ and $b_k\sim\mathcal{N}(0,1)$ are
independent.  The quantity $\rho_1$ ($0\leq\rho_1\leq 1$)
represents the cross-correlation at the AP side, and $1-\rho_1$
represents the cross-correlation at the user side. The spatial
correlation is reflected via the correlation between $a_m$,
$m=1,\ldots, M$, and the correlation between $b_k$, $k=1,\ldots,
K$. We define $\rho_{2,mm'}\triangleq\EX{a_ma_{m'}^\ast}$ and
$\rho_{3,kk'}\triangleq\EX{b_kb_{k'}^\ast}$. These correlation
coefficients depend on the spatial locations of the APs and on the
users:
\begin{align}\label{eq: spatialCorr1}
\rho_{2,mm'}
    =
    2^{-\frac{d_\mathrm{a}(m,m')}{d_{\mathrm{decorr}}}}, \quad \rho_{3,kk'}
    =
    2^{-\frac{d_\mathrm{u}(k,k')  }{d_{\mathrm{decorr}}}},
\end{align}
where $d_\mathrm{a}(m,m')$ is the distance between the $m$th and
$m'$th APs, $d_\mathrm{u}(k,k')$ is the distance between the $k$th
and $k'$th users, and $d_{\mathrm{decorr}}$ is the decorrelation
distance which depends on the environment. This model for the correlation coefficients has been
validated by practical experiments
\cite{WTN:08:VT}.

For the comparison between Cell-Free Massive MIMO and small-cell
systems, we consider the per-user throughput which is defined as:
$$S_k^{\tt A} = B\frac{1-\tau/T}{2}R_k^{\tt A},$$
where ${\tt A} \in \{\text{cf}, \text{sc}\}$ corresponding to
Cell-Free Massive MIMO and small-cell systems respectively, $B$ is
the spectral bandwidth, and $T$ in the length of the coherence
interval in samples. Note that  both Cell-Free Massive MIMO and
small-cell systems need $K$ pilot sequences for the channel
estimation. There is no difference between  Cell-Free Massive MIMO
and small-cell systems in terms of pilot overhead. Furthermore, we
assume that both systems use the same set of pilot sequences. We
choose $T=200$ samples, $B=20$ MHz, $d_{\mathrm{decorr}}=0.1$ km
and $\rho_1=0.5$.

Figure~\ref{fig:5} compares the cumulative distribution of the
throughput per user for small-cell and Cell-Free systems, for
$M=60$, $K=20$, and $\tau=10$, with and without shadow fading
correlation. Compared to the small-cell systems, the throughput of
the Cell-Free systems is much more concentrated around its median.
Without  shadow fading correlation, the 95\%-likely throughput
of the Cell-Free system is about $15$ Mbits/s which is $17$ times
higher than that of the small-cell system (about $0.85$
Mbits/s).  We can see that the small-cell systems are more
affected by shadow fading correlation than the Cell-Free Massive MIMO
systems are.  In this example, the shadow fading correlation
reduces the 95\%-likely throughput by factors of $4$ and
$2$ for small-cell and Cell-Free systems, respectively, compared
to the case of uncorrelated shadowing.

\section{Conclusion} \label{Sec:Conclusion}

We analyzed the performance of Cell-Free Massive MIMO
systems, taking into account  the effects of  channel estimation.
We further compared the performance of  Cell-Free
Massive MIMO to that of  small-cell systems.

The results show that Cell-Free systems
can significantly outperform small-cell systems in terms of throughput.  The
95\%-likely per-user throughput of Cell-Free Massive MIMO is
almost $20$ times  higher than for a small-cell system. Also,
Cell-Free Massive MIMO systems  are more robust to
shadow fading correlation than small-cell systems.


\end{document}